\documentclass[a4paper,12pt]{article}
\usepackage{graphicx}
\usepackage{epstopdf}
\usepackage{amsfonts,amsmath,amssymb}
\usepackage{hyperref}

\usepackage{epsfig,multicol}

\newcommand\fverb{\setbox\pippobox=\hbox\bgroup\verb}
\newcommand\fverbit{\egroup\item[\fbox{\unhbox\pippobox}]}

\newbox\pippobox

\begin{document}
\title{\bf Cosmological Constant Implementing Mach Principle in General Relativity}
\author{ Nadereh Namavarian\thanks{Electronic address: n\_namavarian@sbu.ac.ir}\ \ and
         Mehrdad Farhoudi\thanks{Electronic address: m-farhoudi@sbu.ac.ir}\,\,
\\
{\small Department of Physics, Shahid Beheshti University, G.C.,
        Evin, Tehran 19839, Iran} }
\date{\small September 7, 2016}
 \maketitle
\begin{abstract}
\noindent
 We consider the fact that noticing on the operational
meaning of the physical concepts played an impetus role in the
appearance of general relativity (GR). Thus, we have paid more
attention to the operational definition of the gravitational
coupling constant in this theory as a dimensional constant which
is gained through an experiment. However, as all available
experiments just provide the value of this constant locally, this
coupling constant can operationally be meaningful only in a local
area. Regarding this point, to obtain an extension of GR for the
large scale, we replace it by a conformal invariant model and
then, reduce this model to a theory for the cosmological scale via
breaking down the conformal symmetry through singling out a
specific conformal frame which is characterized by the large scale
characteristics of the universe. Finally, we come to the same
field equations that historically were proposed by Einstein for
the cosmological scale (GR plus the cosmological constant) as the
result of his endeavor for making GR consistent with the Mach
principle. However, we declare that the obtained field equations
in this alternative approach do~not carry the problem of the field
equations proposed by Einstein for being consistent with Mach's
principle (i.e., the existence of de~Sitter solution), and can
also be considered compatible with this principle in the Sciama
view.
\end{abstract}
%
\medskip
 \noindent
 {\small PACS numbers: 04.20.-q; 98.80.Es; 04.90.+e; 04.20.Cv\newline
 }

\vspace*{25pt}
\section {Introduction}
\indent

In the one hundredth anniversary of the advent of GR, it would be
instructive to investigate the intellectual roots of this theory.
Actually, we propose to scrutinize the well--known fact that
Einstein was influenced by Mach's thoughts and positivism,
particularly, the idea that one does~not know the meaning of a
concept unless one has a method for measuring it. Applying this
idea in physics leads to the belief that one is~not allowed to
enter any concept into physics unless an operational definition is
proposed for it. In this respect, Einstein's view, and also
Bridgman's effort~\cite{Bridgman}, gave origin to a method of
philosophical investigation on the meaningfulness of scientific
theories that nowadays goes under the name of {\it operationism},
however, that a theory calls for experiments to prove or disprove
its assertions is certainly not~new in physics. In this regard,
the footprint of this belief can be traced to the appearance of
special relativity~\cite{Woodhouse}. Indeed, Einstein's attempt
for making a consistency between the Newtonian mechanics and the
Maxwell theory led him to provide operational definitions for the
concepts of simultaneity, time interval and spatial interval, that
were finally manifested in the theory of special relativity.
However, although special relativity can be considered as a good
substitution for the Newtonian mechanics, that is also consistent
with the Maxwell equations, but still it suffers from the lack of
an operational definition for the inertial observer.

In this context, the existed operational definition for the
inertial observer, that was proposed by Newton, was unacceptable
for Einstein because it introduces the concept of absolute space
which itself has no operational meaning and has mostly been
considered as a metaphysical one that, usually believed, there is
no room for it in physics. In this respect, Mach's idea about the
inertia could have provided Einstein with an acceptable
operational definition for the inertial observer as an observer
who moves at a constant velocity with respect to the far stars. In
fact, Mach was one of those who were against the absolute space
and absolute motion as pure mental constructs that cannot be
produced in experience~\cite{Mach1977}, and instead, he presumed
that the appearance of the inertial force can be related to the
relative motion with respect to the far massive
bodies~\cite{Mach1977}--\cite{Lichtenegger-2005}. As, in the
earliest expression~\cite{Barbour} of Einstein's realization of
Mach's idea (that can be traced in his paper in
1912~\cite{Einstein1912}), he mentioned that Mach's idea suggests
that the whole inertia of a material point can be considered as
the result of the presence of all other masses depended on a kind
of interaction with them. Such a realization led Einstein to
conclude two principles, namely, the general covariance and the
equivalence principles. Then, his endeavors, in employing these
two principles and implementing the Mach idea through
determination of the metric tensor, $g_{\mu \nu}$, via a tensor
that describes matter and energy, resulted in the manifestation of
the theory of GR~\cite{Einstein-1915,Einstein-1916}. He also hoped
that this theory would~not have any solution in the absence of
matter as the consequence of Mach's principle\rlap.\footnote{More
specific, the {\it strong} version of the Mach principle (namely,
if there is no matter then there is no geometry). However, the
resulted theory only satisfies its {\it weak} version (namely, the
matter distribution determines the geometry).}\ Nevertheless, even
the trivial Minkowski spacetime exists as a vacuum solution of it,
and this fact caused Einstein to be unsatisfied with the theory.

To remedy this issue, at first he thought the problem as the
consequence of boundary conditions at infinity, and thus,
attempted to remove the Minkowski solution by imposing suitable
boundary conditions which do~not impart any absolute structure to
spacetime~\cite{Barbour}. However, he did~not succeed. After a
while, he abandoned his boundary condition formulation of Mach's
principle, and tried to implement Mach's idea through proposing a
model for the universe~\cite{Einstein1917}; wherein, nowadays, it
is believed that a cosmological model depends more on symmetries
than on initial conditions. In this context, Einstein claimed
that, in a finite and closed universe with no boundary region, the
local metric is~not determined in part by the boundary conditions,
but rather only by the matter distribution in the universe. In
other words, his impression was that by having no boundary region,
there would be no room for non--Machian determination of the
metric. In this scenario, he was forced to enter an additional
constant term (which the theory also allows its inclusion) into
the theory, and presented the field equations in the
form\footnote{Throughout this work, the signature is $(-,+,+,+)$,
the lower case Greek indices run from zero to three, and we use
units in which $\hbar = 1= c$.}
\begin{equation}\label{EField}
G_{\mu \nu } - \Lambda {g_{\mu \nu }} =  8\pi G\, T_{\mu \nu }
\end{equation}
in four dimensions, where $G_{\mu \nu}$ is the Einstein tensor as
a function of the metric and its derivatives, $ {T_{\mu \nu }} $
is the energy--momentum tensor, and $\Lambda $ is the cosmological
constant. Now, Einstein expected that these amended equations
would~not admit any matter--free solution in the model of a closed
universe. However, in the early 1917, a vacuum solution was found
by de~Sitter~\cite{deSitter}. Thus, despite his later attempts for
showing the theory to be consistent with Mach's
principle~\cite{Einstein1922}, he finally gave up the struggle
and, his enthusiasm for Mach's principle began to
decrease~\cite{Barbour,Pais}. Nevertheless, the cosmological
constant still survives in the theory and has obtained some other
applications.

By this brief consideration, we have endeavored to emphasize on
the fact that focusing on the operational meaning of physical
concepts could have played an impetus role (at the beginning) in
the appearance of GR\rlap.\footnote{Although, Einstein gave
convincing argument against the absolute time by use of the
operational method, but Bridgman~\cite{Bridgman49} declares his
failure in carrying it over into GR.}\
 Hence, still insisting on
this point of view, let us also consider the operational
definition for the factor $G$ in the Einstein equation. In this
regard, it is well--known that this factor in GR is a constant
which enters into the theory by making it consistent with the
Newtonian theory of gravitation in the weak field limit. Actually,
it is the same gravitational constant that appears in the
Newtonian law of gravitation, and hence, the operational
definition for it is a dimensional constant that can be gained
through the measurement in an appropriate experiment, such as the
Cavendish experiment~\cite{Cavendish}. For a historical review,
and the list of attempts that have been performed for measuring
this constant, see Ref.~\cite{Speake}.

As all these experiments specify the value of this coupling
constant locally (at most in the scale of solar system), one can
say that G, and hence the Einstein equation, is operationally
meaningful only in a local area. Regarding this point, and
insisting on the viewpoint that any concept in a theory should be
operationally meaningful, pose a question on how one can apply
such a theory to the large scale of the universe as a whole. In
this respect, as the conformal invariant theories do~not include
any dimensional ``constant'' and also are free from belonging to
any specific scale, it seems that it would be instructive to
consider GR as a conformal invariant model and then, reduce the
model to a theory for the cosmological scale by breaking down the
conformal symmetry. This is a task that we propose to perform in
this work. Hence, in the next section, we take a glance on the
conformal invariance while emphasizing on our intended points.
Then, in Sect.~3, we introduce the desired model for our purpose,
and conclude the outcomes. Some final remarks are left to the last
section.

\section {Conformal Invariance}
\indent

It is evident that the values of the units of mass, length and
time employed in any theory should be arbitrary. Regarding this
point, the principle of conformal invariance has been developed as
a principle that requires all fundamental equations in physics to
be invariant under local changes of the unit systems
used~\cite{Dicke}--\cite{Faraoni1999}. In principle, these changes
correspond to conformal transformations of the metric which,
loosely speaking, stretch/shrink all lengths and durations via
spacetime--dependent conversion factors. This requirement poses a
fundamental symmetry, namely, the conformal symmetry in physics
which is of ever--increasing attention in modern physical
theories. As in this regard, many interesting results have been
obtained by applying this symmetry, see, e.g.,
Refs.~\cite{20}--\cite{27}.

In addition, in the conformal invariant theories, it is
well--known that in the absence of dimensional parameters, the
conformal invariance requires the vanishing of the trace of the
energy--momentum tensor of the matter conformally coupled to
gravitation, see, e.g., Ref.~\cite{Wald1984}. Also, in such
theories, in the presence of dimensional parameters, the conformal
invariance can be established if such parameters are conformally
transformed according to their dimensions as
well~\cite{Bekenstein}. Indeed, in the conformal invariant
material theories, the dimensional parameters are replaced by
scalar fields to allow for the local changes of the unit systems,
and thus, a general feature of such theories is the presence of
varying dimensional coupling constants. It is also well--known
that breaking down the conformal symmetry can be performed by
introducing a dimensional constant into the theory. Regarding
these points, it is possible to acquire gravitational coupling
through conformal symmetry breaking by looking at the massless
scalar matter field conformally coupled to gravity as a variant of
GR, see Ref.~\cite{19}. Inspired by such an approach, and noticing
that conformal transformations are also considered as
transformations between metrics belonged to different scales, we
propose a method for applying GR to cosmological scale.

Thus, in the next section, first we generalize GR to a conformal
invariant theory via replacement of the gravitational constant by
a scalar field (to let the unit systems being arbitrary at every
point) and choosing suitable coefficients. As conformal
transformations can also be assumed to be transformations between
metrics belonged to different scales, hence, the conformal
symmetry makes the new theory free from belonging to any scale.
Then, we break the conformal symmetry to single out a preferred
conformal frame corresponding to a specific scale. As breaking
down the conformal symmetry also amounts to considering a constant
value for the scalar field (i.e., the varying gravitational
coupling), we perform the symmetry breaking by fixing the value of
the scalar field in a way that the theory is reduced to a
conformal frame which is characterized by the large scale
characteristics of the universe as a whole. Thus, in this way, we
apply GR to cosmological scale.

\section {The Model}
\indent

Following the mentioned explanations, in a $4$--dimensional
spacetime, we replace the Einstein--Hilbert action by the
well--known conformal invariant action functional\footnote{That is
usually used as a model for a matter field conformally coupled to
gravity. This action may also be considered corresponding to the
action of Brans--Dicke theory in the special case of
$\omega=-3/2$, where under transformation $\overline{g}_{\mu \nu }
= \varphi^2 g_{\mu\nu}$, one can get it as
$S[\overline{g}_{\mu\nu}]=\frac{1}{2}\int {d^4}x\sqrt
{-\overline{g}}\left(\frac{1}{6}\overline{R}+\frac{1}{2}\lambda\right)$
without the scalar field~\cite{FFS2010}. Nevertheless, it should
be noted that the scalar field in the Brans--Dicke theory has a
different meaning than the one in our approach. In this theory,
the value of the gravitational constant varies with position as
the consequence of Mach's principle, i.e. the effect of matter
distribution in the universe on the value of this constant at
every point. However, in action~(\ref{action}), $G$ has been
replaced by a scalar field as the consequence of the conformal
invariance and the arbitrariness of unit systems at every point,
actually, the same as the scalar field proposed by Dirac in the
Weyl--Dirac action~\cite{Dirac1973}. }\
\begin{equation}\label{action}
{S[\varphi ,g_{\mu\nu}]} = \frac{1}{2}\int {{d^4}x\sqrt { - g}
\left( \frac{1}{6}R{\varphi^2} +{\partial_\mu }\varphi
{\partial^\mu }\varphi + \frac{1}{2}\lambda {\varphi^4} \right)},
\end{equation}
where $R$ is the Ricci scalar, $\varphi $ is a real scalar field
non--minimally coupled to the gravity (that as explained, it is a
substitution for the gravitational constant in GR), and $\lambda $
is a dimensionless coupling constant. Action~(\ref{action}) is
invariant under the conformal transformations\footnote{In a
spacetime with arbitrary dimension $n$, the conformal invariance
can be achieved if it is accompanied by the appropriate rescaling
of the scalar field as $\varphi \to \Omega^w(x)\varphi$, where the
conformal weight $w$ depends on the dimension of the scalar field
of the model and is $w=1-n/2$, see, e.g., Ref.~\cite{bida}.
However, the resulted two new fields still remain independent. }
\begin{equation}\label{transformations}
\varphi \to \tilde \varphi = {\Omega^{ - 1}}(x)\varphi
\qquad\qquad {\rm and} \qquad\qquad {g_{\mu \nu }}(x) \to {{\tilde
g}_{\mu \nu }}(x) = {\Omega^2}(x){g_{\mu \nu }}(x),
\end{equation}
apart from a complete divergent term that gives no contribution to
the variation of the relevant action, where the conformal factor
$\Omega (x)$ is a non--vanishing arbitrary smooth function of the
spacetime coordinates.

Before adding an action for the matter, let us remind that the
establishment of the conformal symmetry in the vacuum sector of a
gravitational model confronts one with a problem concerning the
incorporation of matter to the gravity. Indeed, as all the
conformal frames are dynamically equivalent, it raises the
question that to which frame the matter should be coupled. As a
choice, we trust the weak equivalence principle (although, it
is~not reliable at quantum level), and according to this
principle, the metric appears in the matter part should be the
same one that describes the gravitational
field\rlap.\footnote{However, in general, one can assume that the
metrics in the gravitational and the matter parts are different
(although conformally related), and achieves interesting results,
see, e.g., Ref.~\cite{27}. }\
 Thus, we consider the total action by adding a matter source ${S_m}$ with
the same metric and independent of $\varphi $ to
action~(\ref{action}), i.e. $S = {S[\varphi ,g_{\mu\nu}]} +
{S_m}$. Now, varying the total action with respect to the scalar
and metric fields yields
\begin{equation}\label{FE for phi}
\Box\,\varphi - \frac{1}{6}R\varphi - \lambda {\varphi^3} = 0
\end{equation}
and, with $\varphi\neq 0$,
\begin{equation}\label{FE for g}
G_{\mu \nu } - \frac{3}{2}\lambda {\varphi^2}{g_{\mu \nu }} =
6{\varphi^{ - 2}}\left(T_{\mu \nu} + T^{[\varphi]}_{\mu
\nu}\right),
\end{equation}
where
\begin{equation}\label{EM of phi}
T^{[\varphi]}_{\mu \nu}\equiv  - {\partial_\mu }\varphi
{\partial_\nu }\varphi + \frac{1}{2}{g_{\mu \nu }}{\partial_\rho
}\varphi {\partial^\rho }\varphi  - \frac{1}{6}\left({g_{\mu \nu
}}\Box\, - {\nabla_\mu }{\nabla_\nu }\right){\varphi^2}
\end{equation}
and as usual the symmetric energy--momentum tensor of matter
is defined as $T_{\mu \nu} \equiv - \left(2/\sqrt{- g}\right)\times\\
\delta {S_m}/\delta {g^{\mu \nu }}$. By taking the trace of
equation~(\ref{FE for g}), while using equation~(\ref{FE for
phi}), one obtains
\begin{equation}\label{traceless EM}
{T^\mu }_\mu  = 0,
\end{equation}
which is similar to the well--known condition on the matter field
conformally coupled to gravitation\rlap.\footnote{Note that, in
four dimensions, with the conformal
transformations~(\ref{transformations}) not only the field
equation~(\ref{FE for phi}) is conformally invariant, but if, in
addition to~(\ref{transformations}), the traceless symmetric
energy--momentum tensor of the matter transforms as ${\tilde
T}_{\mu\nu}=\Omega^{-2}(x)\, T_{\mu\nu}$, then the field
equation~(\ref{FE for g}) and the conservation equation
$\nabla_{\mu}T^{\mu\nu}=0$ are also conformally invariant, i.e.
${\tilde\nabla}_{\mu}{\tilde T}^{\mu\nu}=0$~\cite{Wald1984}.}

At this stage, as explained before, we break the symmetry and fix
the value of the scalar field by introducing a dimensional
constant, say $\Lambda$, into the theory. For this purpose, we add
\begin{equation}\label{lagrange multiplier}
{S_{\alpha (x)}} = \frac{1}{2}\int {{d^4}x\sqrt { - g}\, \alpha
(x) \left(\frac{1}{2}\lambda {\varphi ^4} - \frac{1}{3}\Lambda
{\varphi ^2}\right)}
\end{equation}
to the total action. In this case, the corresponding
equations~(\ref{FE for phi}) and~(\ref{FE for g}) are
\begin{equation}\label{FE for phi 1}
\Box\,\varphi - \frac{1}{6}R\varphi -(1+\alpha) \lambda
{\varphi^3}+\frac{1}{3}\alpha \Lambda \varphi = 0
\end{equation}
and, with $\varphi\neq 0$,
\begin{equation}\label{FE for g 1}
G_{\mu \nu } - 3\left[\frac{1}{2}(1+\alpha)\lambda
{\varphi^2}-\frac{1}{3}\alpha\Lambda\right]{g_{\mu \nu }} =
6{\varphi^{ - 2}}\left(T_{\mu \nu} + T^{[\varphi]}_{\mu
\nu}\right),
\end{equation}
while the variation with respect to $\alpha $ gives a constant
value to the scalar field, i.e. $\varphi^2 = 2\Lambda/(3\lambda)$,
although its value is~not definite yet\rlap.\footnote{This
condition is~not a holonomic constraint, and we have introduced an
auxiliary field $\alpha(x) $ to have such a condition, wherein
with the obtained constant value of $\varphi$, the field equations
also yield $\alpha =9\lambda T^{\mu}{}_{\mu}/(2\Lambda^2)$.}\
 However, by substituting this constant value
of $\varphi$ into the trace of equation~(\ref{FE for g 1}), one
also gets
\begin{equation}\label{varphi}
{\varphi ^2} =  - \frac{{{T^\mu }_\mu }}{{R/6 + 2\Lambda/3 }},
\end{equation}
i.e., as expected, the conformal symmetry breaking is equivalent
to coupling a matter with non--zero trace energy--momentum tensor
to gravity, while singles out a specific conformal frame
corresponding to the constant value of the scalar field.

Now, to consider the symmetry breaking as a cosmological effect,
we plausibly take $\Lambda$ of the order of observational bound on
the cosmological constant, i.e. $R_0^{-2}$, where $R_0$ is the
radius of the universe at the present epoch (i.e., roughly the
Hubble distance). We also approximate ${T^\mu}_\mu$ by its average
value, that is the mass density of the universe in central mass
frame, i.e. $\left\langle {{T^\mu }_\mu } \right\rangle
\sim-{M_0}{R_0}^{ - 3}$ (where the negative sign is due to the
signature), while neglecting the (positive) energy content of the
universe (like photons), as these are identically traceless. By
considering these approximations in relation~(\ref{varphi}), the
Ricci scalar should also have a constant value (just as in GR when
the trace of the energy--momentum tensor is a constant). Further,
following the symmetry breaking as the cosmological effect, and as
the dimension of this constant is one over length square, we take
it to be of the order of $R_0^{-2}$ too, for $R_0$ being a length
that characterizes the size of the universe. Then, replacing back
these approximations into relation~(\ref{varphi}) leads to the
estimation of constant value of $\varphi $ as
\begin{equation}\label{approximation}
{\varphi ^2} \sim\frac{6}{5} \frac{{{M_0}}}{{{R_0}}} \sim {G^{ -
1}},
\end{equation}
where, in the second proportionality, we have used the well--known
empirical cosmological coincidence dimensionless relation
$GM_0/R_0\sim 1$ \cite{Sciama-1953}--\cite{Ray2007}, that is the
Schwarzschild radius of the universe agrees with its
size\rlap.\footnote{It also infers that the inertial energy of a
particle (with mass $m$) is due to the gravitational potential
energy of the matter of the universe upon it, i.e.
$mc^2-GM_0m/R_0\sim 0$, more or less a mathematical formulation of
the Machian point of view~\cite{Sciama,ShojaieFarhoudi}. }\
 At last, by inserting the obtained fixed value of $ \varphi $ into
the field equations~(\ref{FE for g 1}), one obtains\footnote{The
constant value of $\varphi$ and those approximations also give
$\lambda =5/(9M_0R_0)$ and $\alpha =-5/2$. In addition, the
proportionality factor $5$ in equation~(\ref{EFE}) can be
improved. That is, if one considers all the employed
approximations as $\Lambda =a\, R_0^{-2}$, $R=b\, R_0^{-2}$,
$T^{\mu}{}_{\mu}=-c\, M_0 R_0^{-3}$ and $ R_0/M_0=d\, G$, hence,
instead of the proportionality factor $5$, one will get the
equality factor $d(b+4a)$. Furthermore, one can usually consider
$d\sim 1\sim b$ up to the first order, and also uses the relation
$\Lambda =3\, \Omega_{\Lambda}H^2$ while assumes $R_0=\beta\,
H_0^{-1}$ (i.e., $R_0=\beta\, d_H$ at the present epoch). Thus,
with the Planck data~\cite{Ade-2013} for $\Omega_{\Lambda}$ at the
present epoch, if the resulted factor is supposed to be $8\pi$, it
will yield $\beta \simeq 2$. }
\begin{equation}\label{EFE}
G_{\mu \nu }- \Lambda {g_{\mu \nu }} \sim 5\, G\, {T_{\mu \nu }}.
\end{equation}
Therefore, in this alternative approach, wherein the attempt has
been performed to achieve an extension of GR for the cosmological
scale, we have also reached to the same value for $G$ as is
obtained locally, and to the same field equations as Einstein
proposed for this scale through his approach of making GR
consistent with Mach's principle. However, in the proposed model,
the field equations~(\ref{EFE}) have been achieved as the result
of breaking down the conformal symmetry by fixing the value of the
scalar field through introducing the cosmological constant.
Moreover, in the absence of matter, the field equations~(\ref{FE
for phi 1}) and~(\ref{FE for g 1}) yield $\Lambda =0$, however,
see also the following properties.

Also, almost the same result can be obtained via this model with
no potential term ($\lambda\varphi^{4}$) in action~(\ref{action}),
but through a different way of symmetry breaking as has been
considered in Ref.~\cite{19}\rlap.\footnote{Note that, there is no
explanation in this paper about the meaning one can attribute to
the scalar field; and however, as mentioned, the approach of this
work inspired us in figuring out the way of applying conformal
symmetry, and then, breaking it as a method for acquiring a
generalization of a theory operationally valid in some scale, for
some other scales.}\
 Actually, the variation of this new action (used in that paper) gives the same field
equations~(\ref{FE for phi}) and~(\ref{FE for g}) with $\lambda=0$
terms, and yet with the same traceless energy--momentum tensor,
i.e. equation~(\ref{traceless EM}). However, to break down the
conformal symmetry in this potential--free action, a dimensional
mass term for the scalar field, i.e. $M^2\varphi^2/2$, has been
added to its related Lagrangian, which leads to the corresponding
field equations
\begin{equation}\label{D FE for phi 1}
{\partial_\mu }{\partial^\mu}\varphi - \frac{1}{6}R\varphi -
{M^2}\varphi = 0
\end{equation}
and
\begin{equation}\label{D FE for g 1}
G_{\mu \nu } - 3{M^2}{g_{\mu \nu }}= 6{\varphi^{ - 2}}\left(T_{\mu
\nu} + T^{[\varphi]}_{\mu \nu}\right),
\end{equation}
with ${T^\mu }_\mu = -{M^2}{\varphi^2}$. Thus, by the same
approximation for the energy--momentum tensor and assuming the
mass $M$ of the order of $R_0^{-1}$, the scalar field is fixed.
Then, inserting it into equation~(\ref{D FE for g 1}), leads to
the same field equations~(\ref{EFE}) (where the factor $5$ has
been replaced by $6$) with an effective cosmological constant, of
the right sign, i.e. $3{M^2}\sim R_0^{-2}$, that in turn, is
proportional to $T^{\mu}{}_{\mu}\mid_{\varphi_{_{\rm fixed}}}$.

Eventually, the field equations~(\ref{EFE}), that are obtained
from this alternative approach, possess the following properties.
\begin{itemize}
\item First, the coupling constant $G$, that is acquired
      via breaking down the conformal symmetry, is related to the mass
      distribution in the visible universe in the form
      of $GM_0/R_0\sim 1$.

\item Second, as the breaking down of the symmetry is
      equivalent to considering a matter field with non--zero trace, hence, the
      existence of matter in the universe is a presupposition for
      obtaining these field equations.

\item Third, the presence of the cosmological constant in the field
      equations is actually the indication of the above presupposition.
\end{itemize}
These properties make the field equations~(\ref{EFE}) different
from those proposed by Einstein.

Now, as the motivation behind the Einstein field equations was
making consistency between GR and Mach's principle, let us
investigate the consistency of the field equations~(\ref{EFE})
with this principle too. In this context, according to the second
and third properties, the vacuum solution for these field
equations is meaningless. Although, by having such a presumption
one cannot check the consistency of these field equations with
Mach's principle (through the vacuum solution) in a closed
universe, but, in this alternative approach, the de~Sitter
solution cannot either be as an evidence against Einstein's
expected consistency between these field equations and the Mach
principle. On the other hand, as the relation between $G$ and the
mass distribution in the visible universe (i.e., $GM_0/R_0\sim
1$), besides being an observational relation, is also a
representation of Mach's principle proposed by
Sciama~\cite{Sciama}, the first property indicates the consistency
of the field equations~(\ref{EFE}) with Mach's principle in the
Sciama view. Therefore, one can claim that in this work, the
cosmological constant is playing the same role as Einstein hoped,
i.e. the cosmological constant introduced into GR has implemented
Mach's principle via this alternative approach.

Also, an accepted view toward Mach's principle is the formulation
by Bondi, i.e. ``local inertial frames are determined through the
distributions of energy and momentum in the universe by some
weighted averages of the apparent motions"~\cite{Bondi}, that has
been illustrated in the cosmological perturbation theory through
the existence of the gauges in the theory called ``Machian
gauges'' in which the distribution of $\delta T^\mu{}_\nu$
determines uniquely and instantaneously the rotations and
accelerations of local inertial frames via Einstein's field
equations~\cite{Bicak}. Moreover, it has been shown that in such a
gauge, only differences of rotation rates of inertial frames are
determinable for the case of a closed universe, and hence, no
absolute rotations exist~\cite{Lynden} in accord with Mach's ideas
that all motions are relative. However, in the field
equations~(\ref{EFE}), as the value of the gravitational coupling
depends on the energy--momentum tensor of the matter in the
universe, it may change the results, and this matter can be
investigated.

On the sidelines, although significant on its own, we should
emphasize that the result of the approach (that implements the
cosmological constant should actually be related to the matter)
somehow reminds one of the interpretation of the vacuum fluid and
the vacuum energy density, see, e.g.,
Refs.~\cite{Barrow2011,WeinbergBook}. And more important, as the
gravitational field equations in the alternative approach cannot
be considered without a non--zero cosmological constant term, and
wherein also, in the reverse side, the matter being somehow
dependent on it, such a situation reminds one of the ether issue,
see, e.g., Ref.~\cite{FarhoudiY} and references therein. In this
respect, we should mention that even Einstein himself
claimed~\cite{Janssen2005,ref27-4} that, in principal, the general
theory of relativity is merely an ether theory.

\section {Final Remarks}
\indent

By insisting on the viewpoint that any concept in a theory should
possess an operational meaning, we have applied GR to the
cosmological scale through its replacement with a conformal
invariant model and then, have reduced this model to a conformal
frame characterized by the large scale characteristics of the
universe. Finally, we have reached to the same field equations (GR
plus a cosmological constant) that Einstein proposed for this
scale with the purpose of making consistency between GR and Mach's
principle, although he was~not successful. However, in this
alternative approach, the theory itself determines the constant
$G$ in the cosmological scale. Also, we have declared that the
field equations obtained in this approach do~not carry the problem
of Einstein's field equations for being consistent with Mach's
principle and, they can be considered compatible with this
principle in the Sciama view as well.

\section*{Acknowledgements}
\indent

We thank the Research Office of Shahid Beheshti University for the
financial support.

\end{document}